\documentclass[aps,prl,superscriptaddress,reprint,longbibliography,floatfix]{revtex4-2} 

\usepackage{tabularx}
\usepackage{multirow}
\usepackage{amsmath,amssymb}
\usepackage{lmodern}
\usepackage{graphicx}
\usepackage{epstopdf}
\usepackage[colorlinks=true,linkcolor=red,bookmarks=true,citecolor=blue,urlcolor=blue]{hyperref}
\usepackage[T1]{fontenc}
\usepackage{color}
\usepackage[separate-uncertainty=true]{siunitx}
\usepackage{lipsum}
\usepackage[plain]{fancyref}
\usepackage{soul}
\usepackage{braket}

\begin{document}
\title{Quantum interference visibility spectroscopy in two-color photoemission from tungsten needle tips}

\author{Ang Li}
\email[]{ang.li@fau.de}
\affiliation{Department of Physics, Friedrich-Alexander Universit\" at Erlangen-N\" urnberg (FAU), \\ Staudtstra\ss e 1, 91058 Erlangen, Germany}

\author{Yiming Pan}
\affiliation{Physics Department and Solid-State Institute, Technion, \\ Haifa 32000, Israel}

\author{Philip Dienstbier}
\affiliation{Department of Physics, Friedrich-Alexander Universit\" at Erlangen-N\" urnberg (FAU), \\ Staudtstra\ss e 1, 91058 Erlangen, Germany}

\author{Peter Hommelhoff}
\email[]{peter.hommelhoff@fau.de}
\affiliation{Department of Physics, Friedrich-Alexander Universit\" at Erlangen-N\" urnberg (FAU), \\ Staudtstra\ss e 1, 91058 Erlangen, Germany}

\date{\today}

\begin{abstract}
When two-color femtosecond laser pulses interact with matter, electrons can be emitted through various multiphoton excitation pathways. Quantum interference between these pathways gives rise to a strong oscillation of the photoemitted electron current, experimentally characterized by its visibility. In this work, we demonstrate two-color visibility spectroscopy of multi-photon photoemission from a solid-state nanoemitter. We investigate the quantum pathway interference visibility over an almost octave-spanning wavelength range of the fundamental ($\omega$) femtosecond laser pulses and their second-harmonic (2$\omega$). The photoemission shows a high visibility of 90\% $\pm$ 5\%, with a remarkably constant distribution. Furthermore, by varying the relative intensity ratio of the two colors, we find that we can vary the visibility between 0 and close to 100\%. A simple but highly insightful theoretical model allows us to explain all observations, with excellent quantitative agreements. We expect this work to be universal to all kinds of photo-driven quantum interference, including quantum control in physics, chemistry and quantum engineering. 
\end{abstract}


\maketitle

\newpage
Manipulating atomic and molecular processes using coherent light lies at the core of quantum control. It can be achieved by tuning quantum interference between two competing pathways by varying the relative phase and amplitudes of bichromatic laser fields, driving the relevant transitions. The resulting quantum mechanical pathway-interference influences the yield of the final states \cite{Shapiro2011, Shapiro1986, Brumer1986, Shapiro1988, Chan1991} and has been first observed in actively controlled photodissociation of molecules \cite{Lu1992,Zhu1995,Charron1995}. Later, this method has been extended to the ionization of atoms, where the energy and angular distributions of emitted electrons are manipulated by tuning the laser parameters \cite{Chen1990, Chan1990, Muller1990, Schumacher1994, Ehlotzky2001}. Also, coherent control of atomic phenomena has been reported in autoionization \cite{Nakajima1993, Nakajima1994, Xenakis1999} and dissociative ionization \cite{Sheehy1995, Thompson1997, Bandrauk2002, Ohmura2004}. Using the second laser field to introduce asymmetry enables the study of the manipulation of high harmonic generation (HHG) \cite{Kim2005, Mauritsson2006, Brugnera2011}, making ultracold molecules \cite{Koch2006}, measuring and controlling tunneling processes \cite{Dudovich2006, Shafir2012}, plasmonic field distribution \cite{Ji2018} and photoelectron holography \cite{Porat2018}. 

This control scheme was also applied to metallic nanoemitters. Photoemission from the surface of metallic needle tips driven by two-color femtosecond laser fields has shown a remarkably large coherence, indicated by a visibility reaching 97.5\% \cite{ForsterPRL,Paschen2017,Cheng-WeiHuang2017,Luo2018,Luo2019}. This photoemission displayed a homogeneous modulation for all emitted electron energies, driven with a two-color frequency pair of fundamental and second-harmonic ($\omega$-2$\omega$). Here, we demonstrate \textit{visibility spectroscopy} of multi-photon photoemission i.e., the visibility of the coherent signal is investigated in terms of the relative phase, intensities and frequencies of the two laser fields ($\omega$, $2\omega$). We find that, over an almost octave-spanning range of tuned wavelengths, that the visibility shows a nearly constant value on the level of 90\% $\pm$ 5\%. We propose a quantum pathway model that attributes the robust interference to an exact substitution of one ($2\omega$)-photon for two $\omega$-photons from two competing pathways. Furthermore, this model explains the observed intensity scaling and visibility with excellent quantitatively agreement over the entire parameter space. 

\begin{figure}[!ht]
\centering
\includegraphics[width=\columnwidth]{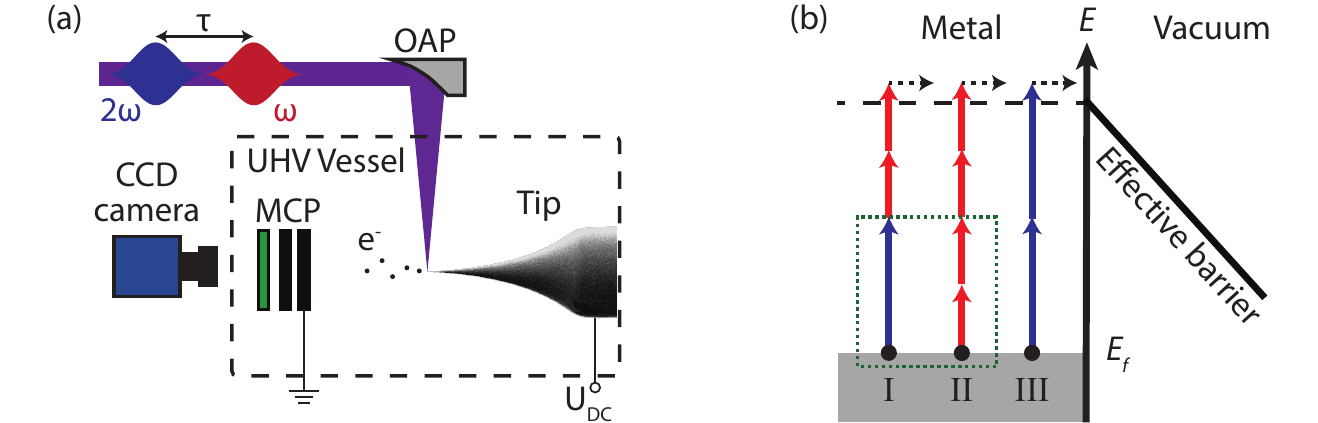}
\caption{\label{fig:setup} Experimental setup and multiphoton absorption in a metallic needle tip.  (a) Femtosecond laser pulses ($\omega$) and their second harmonic ($2\omega$) are focused onto a tungsten needle tip with an off-axis parabola (OAP). Photoelectrons from the tip biased with $U_{DC}$ are detected on a micro-channel plate (MCP) with a phosphor screen and are imaged on a CCD camera. (b) Schematic energy diagram of photoemission processes through three pathways (I, II, III). The green dotted frame marks the substitution of one ($2\omega$)-photon for two $\omega$-photons in the absorption pathways. }
\end{figure}

Our experimental setup is schematically depicted in Fig.~\ref{fig:setup} (a). Ultrashort $\sim$\SI{67}{\femto\second} laser pulses at (angular) center frequency $\omega$ and their second harmonic pulses at $2\omega$ are focused onto a [110]-oriented tungsten needle tip, which is etched electrochemically from a tungsten wire to a sharp tip. Field ion microscopy yields a tip apex radius of \SI[separate-uncertainty=true]{14(1)}{\nano\meter} \cite{Muller1965}. The second harmonic pulses are phase stable relative to the fundamental pulses due to the parametric $2\omega$ generation process and are temporally delayed by a variable time delay $\tau$. The two-color laser pulses are focused  on the tip with a \SI{152}{\milli\meter} effective focal length \SI{90}{\degree} off-axis parabolic mirror situated outside of the ultra-high vacuum (UHV) vessel. Inside the UHV vessel, photoelectrons from the needle tip are detected by a microchannel plate (MCP) detector with a phosphor screen, which is imaged by a CCD camera. 

The multi-photon processes at the surface of the needle tip driven by the two-color laser field are schematically presented in Fig.~\ref{fig:setup} (b). Electrons inside the metal surface of the needle tip can be excited to an energy of $E_{\mathrm{f}}+4\hbar\omega$ through three photon-induced pathways, and overcome the effective barrier. Interference between all three pathways shown should, in theory, be observed, resulting in photoemission oscillations depending on the relative phase between the $\omega$ and $2\omega$ pulses. Particularly, a substitution of one ($2\omega$)-photon for two $\omega$-photons from the interfering pathways I and II (green dotted frame in Fig.~\ref{fig:setup} (b)) should lead to an oscillation with a frequency of $2\omega$, whereas the interference between the pathways II and III should result in the doubled oscillation frequency of $4\omega$ \cite{Shapiro2011}. We note that we observe oscillations with $2\omega$ only, thus we may neglect the interference between the pathways II and III, as shown and discussed below. Furthermore, interference between the pathways I and III is essentially equivalent to the interference between the pathways I and II, since both of them feature the substitution of one ($2\omega$)-photon for two $\omega$-photons, hence the following discussion is focused on the interference between the pathways I and II. 

\begin{figure*}[!ht]
\centering
\includegraphics[width=\textwidth]{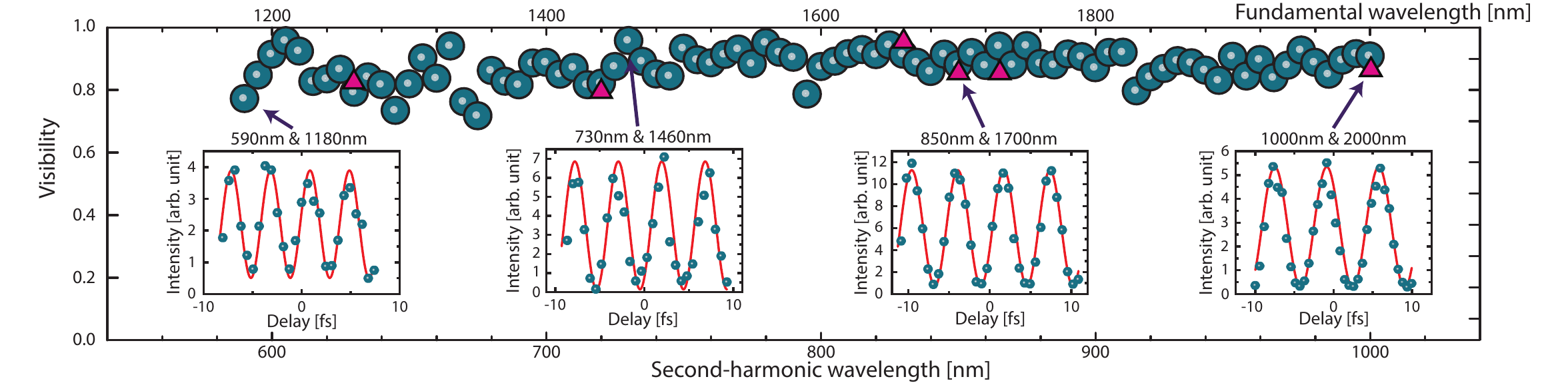}
\caption{\label{fig:spectrogram} Visibility of the two-color coherent control in multi-photon photoemission. Visibilities of the photocurrent oscillations are plotted as a function of the corresponding second-harmonic wavelength of the applied wavelength pairs, which are varied from \SI{590}{\nano\meter} ($2\omega$) and \SI{1180}{\nano\meter} ($\omega$) to \SI{1000}{\nano\meter} ($2\omega$) and \SI{2000}{\nano\meter} ($\omega$). A high visibility of 90\%  $\pm$ 5\% with a roughly constant distribution is found. Visibilities calculated from the theory (Eq.~\eqref{visibility2}) are shown as pink triangles. Insets: Typical photocurrent oscillations when the needle tip is illuminated by the two-color laser field. Photocurrent intensities detected on the MCP detector screen are shown depending on the time delay $\tau$ between $\omega$ and $2\omega$ pulses. Sine fits to the data are displayed as red solid lines, clearly revealing an oscillation frequency of 2$\omega$. }
\end{figure*}

We observe strong oscillations of the photocurrent depending on the time delay $\tau$ when the $\omega$ and $2\omega$ pulses are close to perfect temporal overlap (insets of Fig.~\ref{fig:spectrogram}). These oscillations of the photocurrent can be well fitted by sinusoidal functions (red solid lines). The frequency of the oscillation obtained from the fitted sinusoidal functions and from the Fourier transform of the measured data corresponds to the second harmonic frequency $2\omega$ to within 4\%, whereas the 4$\omega$ component is not visible and is suppressed by at least \SI{-12.1}{\dB}. From the sinusoids, we obtain the visibility of the photocurrent oscillations, defined as 
\begin{equation}
\label{VisDefini}
    V = \frac{N_{\mathrm{max}}-N_{\mathrm{min}}}{N_{\mathrm{max}}+N_{\mathrm{min}}} , 
\end{equation}
where $N_{\mathrm{max}}$ corresponds to the maximum and $N_{\mathrm{min}}$ to the minimum of the sinusoid. Visibilities of these oscillating photocurrents as a function of wavelength are presented in Fig.~\ref{fig:spectrogram}. They reach up to 96\% at the wavelength pairs (\SI{1210}{\nano\meter} \& \SI{605}{\nano\meter}) and (\SI{1460}{\nano\meter} \& \SI{730}{\nano\meter}). Clearly, at each wavelength pair within the large wavelength range investigated (from \SI{1180}{\nano\meter} to \SI{2000}{\nano\meter} in steps of \SI{10}{\nano\meter}), we observe strong oscillations in the photocurrent. Significantly, a nearly constant distribution of the visibility with an average of 90\% $\pm$ 5\% is observed. Within the measured wavelength range, the visibilities are always larger than 72\%, indicating a robust and wavelength-independent interference process. 

To obtain a single data point in Fig.~\ref{fig:spectrogram}, we have recorded short $\omega$ - $2\omega$ delay spectra like shown in the insets of Fig.~\ref{fig:spectrogram}. We have chosen intensities such that we always operate under stable conditions. The plotted visibilities are achieved by setting $I_{\mathrm{\omega}}$ and varying $I_{\mathrm{2\omega}}$  to obtain the maximum visibility. We find that the maximum visibility intensity ratio $I_{\mathrm{2\omega}}/I_{\mathrm{\omega}}$ equals 93\% $\pm$ 31\% for all data points, at an incident fundamental intensity of \SI{7 \pm 1.1e10}{\watt\per\centi\meter\squared}. 

To elucidate the underlying physical process, we propose a simple yet insightful theoretical model based on multi-photon absorption physics from the metal needle tip. We assume that the interaction between the bound electron inside of the metal and the laser field can be described by the Hamiltonian $H_{\mathrm{int}} = -(p\cdot A) e/m $ in the minimal coupling approximation, with electron charge $e$, effective mass $m$, momentum $p$ and the vector potential $A$ describing the nearfield-enhanced laser field. Here, the metallic ground state is filled by Bloch electrons with energy $E$, beneath the Fermi energy $E_{\mathrm{F}}$. The initial electronic state, therefore, takes the form $\ket{\Phi_{\mathrm{i}}(t)}=\exp\left(-iEt/\hbar\right)\ket{k}$ with the electron wavenumber $k \leq k_{\mathrm{F}}$ (Fermi wavenumber). The enhanced laser field is approximately given by $A=g_\omega F \sin(\omega t + \phi_0)$, where, $g_{\mathrm{\omega}}=-\xi_{\mathrm{\omega}}/\omega$ depends on the dimensionless field enhancement factor $\xi_{\mathrm{\omega}}$, which can be determined through numerical simulations \cite{Thomas2015}. The parameter $F$ is the amplitude of the incident laser electric field and $\phi_0=\phi_{\mathrm{\omega}} -\phi_{\mathrm{e}}$ is the phase difference between the laser pulse ($\phi_{\mathrm{\omega}}$) and the phase of the Bloch electron ($\phi_{\mathrm{e}}$). 

By solving the time-dependent Schr\" odinger equation in the interaction picture, we obtain the final state of the electron after the interaction,  
\begin{equation}
\label{FinalStateSingle}
    \ket{\Phi_{\mathrm{f}}(t)}=\sum_{n=-\infty}^{\infty} J_{\mathrm{n}}\left( \frac{e k \Delta t g_{\mathrm{\omega}}}{m} F \right) e^{in(\omega t+\phi_0)} \ket{\Phi_{\mathrm{i}}(t)} ,
\end{equation}
where $J_{\mathrm{n}}$ are the $n$th-order Bessel functions of the first kind, $n$ corresponds to the order of the multi-photon process. Specifically, the order $n$ takes integer values ($n$ = 0, $\pm$1, $\pm$2 ...), denoting photon absorption ($n$ < 0) and photon emission ($n$ > 0) with photon energy of $\hbar\omega$. $\Delta t$ is the duration of the interaction between the electron and the enhanced laser field, which is approximated as $\Delta t \approx \SI{200}{\atto\second}$, based on the length scale (few angstroms \cite{Neppl2015}) of the interaction region between the screened electromagnetic field and the Bloch electron with Fermi velocity $v_{\mathrm{F}}$ extending into vacuum. The dimensionless argument of the Bessel function describes the effective exchanged photon number $\eta_{\mathrm{\omega}}= \frac{e k \Delta t g_{\mathrm{\omega}}}{m} F$. For the multi-photon photoemission process through pathways I and II in Fig.~\ref{fig:setup} (b), the final electronic state is composed of the terms $J_{\mathrm{-1}} (\eta_{\mathrm{2\omega}}) e^{-(2\omega t + \phi_{\mathrm{2\omega}}-\phi_{\mathrm{e})}}$, $J_{\mathrm{-2}}(\eta_{\mathrm{\omega}})e^{-2(\omega t +\phi_{\mathrm{\omega}}-\phi_{\mathrm{e}})}$ and $J_{\mathrm{-4}}(\eta_{\mathrm{\omega}})e^{-4(\omega t +\phi_{\mathrm{\omega}}-\phi_{\mathrm{e}})}$. The resultant final state population via pathways I and II then reads:  
\begin{widetext}
\begin{equation}
\label{TransitionRate}
\begin{aligned}
    \left| \Phi_{\mathrm{f}} (\omega,2\omega) \right| ^2  =  &\left| J_{\mathrm{-4}}\mathrm{(\eta_{\mathrm{\omega}})}\right|^2 +\left| J_{\mathrm{-1}}\mathrm{(\eta_{\mathrm{2\omega}})} J_{\mathrm{-2}}\mathrm{(\eta_{\mathrm{\omega}})} \right|^2 \\ & + 2\left| J_{\mathrm{-4}}\mathrm{(\eta_{\mathrm{\omega}})} J_{\mathrm{-1}}\mathrm{(\eta_{\mathrm{2\omega}})} J_{\mathrm{-2}}\mathrm{(\eta_{\mathrm{\omega}})}\right|  \cos{\left( \phi_{\mathrm{2\omega}}-2\phi_{\mathrm{\omega}}+\phi_{\mathrm{e}} \right)} , 
\end{aligned}
\end{equation}
\end{widetext}
where the first two terms denote the contributions from pathways I and II separately, and the third term describes the interference between the pathways, oscillating with the relative phase $\phi_{\mathrm{2\omega}}-2\phi_{\mathrm{\omega}}+\phi_{\mathrm{e}}$. Using the asymptotic form of the Bessel function of the $n$-th order $J_{\mathrm{n}}(\eta_{\mathrm{\omega}})=(-2)^n/(-n!\cdot\eta_{\mathrm{\omega}}^n)$, Eq.~\eqref{TransitionRate} can be calculated explicitly. Together with Eq.~\eqref{VisDefini} we obtain the visibility as,  
\begin{align}
\label{visibility1}
      V  & = \frac{2\alpha I_{\mathrm{\omega}}\sqrt{\beta I_{\mathrm{2\omega}}}}{\alpha^2 I_{\mathrm{\omega}}^2 + \beta I_{\mathrm{2\omega}}} , 
      \\ 
\label{visibility2}
      & = \frac{2 \sqrt{\Gamma \Tilde{I}}}{1+\Gamma\Tilde{I}} , 
\end{align}
where $\alpha$ and $\beta$ are the proportionality factors for the absorption of $\omega$ and $2\omega$ photons, respectively. By substituting $\beta I_{\mathrm{2\omega}}/\alpha^2 I_{\mathrm{\omega}}^2$ with $\Gamma \Tilde{I}$, the expression of the visibility in Eq.~\eqref{visibility1} is reduced to a simple equation [Eq.~\eqref{visibility2}]. The dimensionless product $\Gamma \Tilde{I}$ describes the branching ratio between the two sub-pathways in I and II. This ratio fully controls the visibility of the quantum pathway interference, independent of the total photon absorption number $n$. A visibility of 100\% will occur if the two sub-pathways contribute equally, i.e., $\Gamma \Tilde{I}=1$. Further, we deduce that the visibility can be tuned by variation of the \textit{intensity} branching ratio $\Tilde{I}=I_{\mathrm{2\omega}}/I_{\mathrm{\omega}}^2$ or the \textit{material} branching ratio $\Gamma=\beta/\alpha^2$, which is related to the optical response of the material. 

\begin{figure*}[!ht]
\centering
\includegraphics[width=\textwidth]{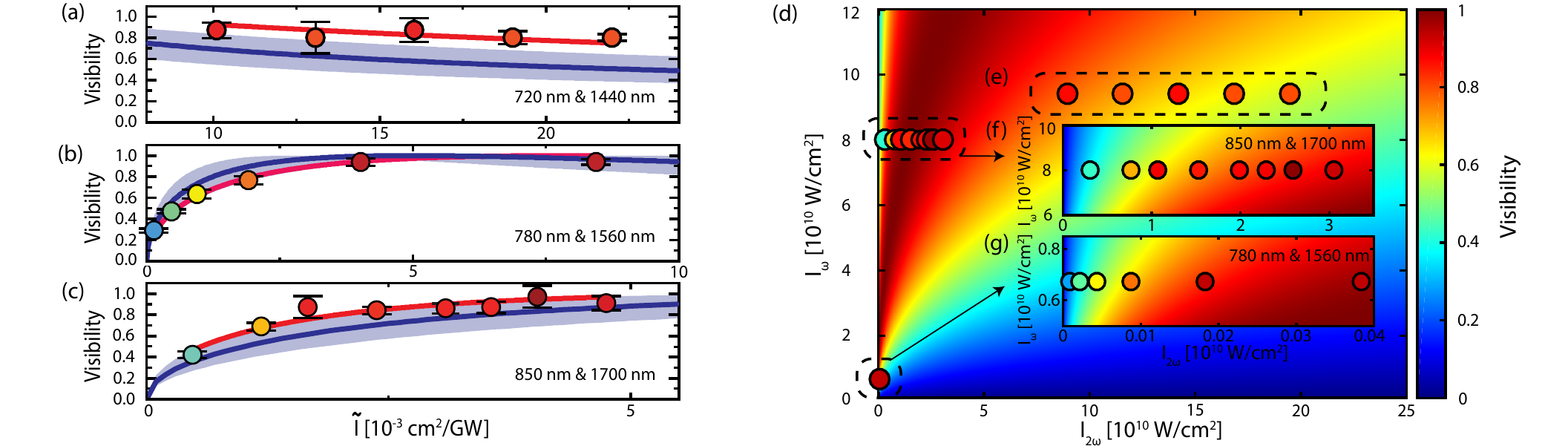}
\caption{\label{fig:theory} Comparison between theory and observations of the visibility scaling as a function of the laser intensities. (a) - (c) Visibility scaling depending on the intensity branching ratio $\Tilde{I}=I_{\mathrm{2\omega}}/I_{\mathrm{\omega}}^2$, measured by varying the second-harmonic intensity $I_{\mathrm{2\omega}}$. The data in (b) are taken from \cite{ForsterPRL}. The color coding of the circles also represents the measured visibilities corresponding to the color bar in (d). Red solid lines in (a) - (c) depict fit curves using Eq.~\eqref{visibility2}, whereas the blue solid lines show the calculated visibility using Eq.~\eqref{visibility2} and Eq.~\eqref{BranchingRatio}. The blue band indicated the uncertainty of the calculated visibility obtained from the 10\% uncertainty assumed in the field enhancement factors. (d) The calculated visibility as a function of fundamental intensity $I_{\mathrm{\omega}}$ and second-harmonic intensity $I_{\mathrm{2\omega}}$ using Eq.~\eqref{visibility1} for the wavelength pair of \SI{720}{\nano\meter} \& \SI{1440}{\nano\meter} (color map). The measured visibilities of (a) - (c) are plotted as colored circles in (d) (black dashed frames) as well as in (f) and (g). (e) The data points show a consistently larger visibility and hence deviate from the theory color map (background color) like they do in (a). (f) and (g) Two zoom-in insets for better visualization, where the background color in (f) is calculated for \SI{850}{\nano\meter} \& \SI{1700}{\nano\meter}, while the background color in (g) is calculated for \SI{780}{\nano\meter} \& \SI{1560}{\nano\meter}, with experimental data points (colored circles) same as shown in (b) and (c). We note the excellent agreement of experimentally measured visibility (circles) and the theoretical one (background color). }
\end{figure*}

Fig.~\ref{fig:theory} (a) – (c) show a measurement of the visibility scaling as a function of the intensity branching ratio $\Tilde{I}$ for three wavelength pairs. The fundamental intensities are as different as $I_{\mathrm{1440nm}}=\SI{9.4e10}{\watt\per\centi\meter\squared}$, $I_{\mathrm{1560nm}}=\SI{6.7e9}{\watt\per\centi\meter\squared}$ and $I_{\mathrm{1700nm}}=\SI{8.0e10}{\watt\per\centi\meter\squared}$. Red solid lines in Fig.~\ref{fig:theory} (a) - (c) are the curves fitted by Eq.~\eqref{visibility2}. Although the measured visibilities follow rather rapid (Fig.~\ref{fig:theory} (b), (c)) or more gentle scaling behaviors (Fig.~\ref{fig:theory} (a)), the analytical expression of the visibility agrees well with all three measurements, confirming the theoretically deduced visibility behavior. 

The material branching ratio $\Gamma$, that jointly controls the visibility with $\Tilde{I}$, can be derived with the effective exchanged photon number $ \eta_{\mathrm{\omega}}=e k \Delta t g_{\mathrm{\omega}} F/m $ as, 
\begin{equation}
\label{BranchingRatio}
    \Gamma = \frac{\beta}{\alpha^2}=\frac{72 m^2 \epsilon_0 c}{ e^2 k_{\mathrm{F}}^2 \Delta t^2} \cdot \frac{\omega^2 \xi_{\mathrm{2\omega}}^2}{\xi_{\mathrm{\omega}}^4} , 
\end{equation}
where $m$ is the electron mass, $\epsilon_0$ is the vacuum permittivity, $c$ is the speed of light in vacuum, $k_{F}=\sqrt{2 m E_f}/\hbar=\SI{1.4e10}{\per\meter}$ is the Fermi wavenumber \cite{Mattheiss1965}. The field enhancement factors $\xi_{\mathrm{\omega,2\omega}}$ are determined by a finite difference time domain (FDTD) simulation for the different wavelengths. This way we can calculate $\Gamma$ (see Supplemental Material for details \cite{supplMat}) and hence obtain the visibilties for different wavelength from the model (pink triangles in Fig.~\ref{fig:spectrogram}). 

Visibilities determined with the calculated $\Gamma$ are shown as blue solid lines in Fig.~\ref{fig:theory} (a) – (c). The blue shadowed areas depict the uncertainty of the visibility due to an assumed uncertainty of the simulated field enhancement factors of 10\%. These results, obtained solely from our model, are also found in good agreement with the experimental results and further confirm the visibility tuning, also with $\Gamma$. Thus, the branching ratio $\Gamma\Tilde{I}$, implying the substitution of one ($2\omega$)-photon for two $\omega$-photons, fully characterizes the visibility [Eq.~\eqref{visibility2}]. 

The visibility is also determined via Eq.~\eqref{visibility1} with the calculated proportionality factors $\alpha$ and $\beta$, as a function of the incident laser intensities $I_{\mathrm{\omega}}$ and $I_{\mathrm{2\omega}}$. The resulting visibility distribution is shown as the background color maps of Fig. \ref{fig:theory} (d), (f) and (g), for the wavelength pairs presented in (a), (c) and (b), respectively. In (d), The calculated value of the proportionality factors $\alpha=\SI{3.15e12}{\centi\meter\squared\per\watt}$ and $\beta=\SI{6.09e12}{\centi\meter\squared\per\watt}$ for \SI{1440}{\nano\meter} and \SI{720}{\nano\meter} are adapted. Clearly, we find a large high-visibility area ($V\geqslant 90\%$) with increasing fundamental intensity $I_{\mathrm{\omega}}$, which is also in agreement with previous theoretical work \cite{Luo2018}. This elegantly explains the large differences in the scaling behavior presented in (a) - (c): These measurements are conducted with different levels of $I_{\mathrm{\omega}}$. In most regions, $\Gamma \Tilde{I}  \gg 1$, hence from Eq.~\eqref{visibility2}, $V \propto \sqrt{\Tilde{I}} \propto I_{\mathrm{\omega}}^{-1}I_{\mathrm{2\omega}}^{1/2}$. Thus, to keep $V$ constant, $I_{\mathrm{2\omega}}$ needs to scale quadratically with $I_{\mathrm{\omega}}$. This notion becomes obvious when insets (f) and (g) are compared: $I_{\mathrm{2\omega}}$ for the maximum visibility is approximately 167=12.9$^2$ times larger in (f) as in (g), whereas the $I_{\mathrm{\omega}}$ is 12 ($\approx 12.9$) times larger, confirming the $V$ scaling behavior. We note that the range of high visibility may become smaller when the field of the fundamental laser field is entering the strong-field regime \cite{Luo2018}. 

In addition, at an appropriate fundamental intensity $I_{\mathrm{\omega}}$, a broad range of $I_{\mathrm{2\omega}}$ is expected to yield a high visibility. For the second-harmonic intensity ratio $I_{\mathrm{2\omega}}/I_{\mathrm{\omega}}=93\% \pm 31\%$ with $I_{\mathrm{\omega}} = \SI{7 \pm 1.1e10}{\watt\per\centi\meter\squared}$ employed in Fig.~\ref{fig:spectrogram}, an average visibility of 87\% with a deviation as small as 7\% is predicted by Eq.~\eqref{visibility1}, in great accordance with the measured visibility of 90\% $\pm$ 5\%. Conversely, for the fundamental intensity $I_{\mathrm{\omega}} = \SI{7 \pm 1.1e10}{\watt\per\centi\meter\squared}$ used in the experiment, an extensive range of the second-harmonic intensity ratios from 12\% to 77\% is predicted to obtain the measured average visibility of 90\%, which is also consistent with the experiment. This further explains the almost constant distribution of the visibilities, measured against a wide spectrum of wavelengths and laser intensity ratios. These additional agreements further support the proposed model, which not only fully explains the observed features in the experimental results, but also provides new insight and further degrees of control in quantum pathway interference. 

Until now, the discussion was based on an estimated interaction duration $\Delta t=\SI{200}{\atto\second}$, resulting from the duration of the Bloch electron, with Fermi velocity $v_{\mathrm{F}} = \SI{1.6e6}{\meter\per\second}$ for tungsten \cite{Mattheiss1965}, traversing the screening length inside the metal and the Bloch wave extension in the vacuum ($\sim$ 3 {\AA} \cite{Neppl2015}). This estimation holds for any laser frequency as long as the fundamental frequency is lower than the plasma frequency of the material (otherwise the screening length becomes much larger) \cite{Bovensiepen2010}. Furthermore, we expect this estimation to also hold for atoms, where the extension of the wave function in vacuum dominates the interaction duration $\Delta t$, which can be varied by changing the height of the ionisation potential. Additionally, the field enhancement factors are determined experimentally in another ongoing work \cite{PhilipTimo}, where the field enhancement factors are directly evaluated from the scaling of the cut-off energy in strong-field electron energy spectra. The experimentally obtained field enhancement factors $\xi_{\mathrm{1560nm}}=5.5 \pm 0.8$ and $\xi_{780nm}=2.8 \pm 0.4$, together with $\Gamma_{\mathrm{780nm}}=$ \SI{1.3e11}{\watt\per\centi\meter\squared} from the fit curve in Fig.~\ref{fig:theory} (c) yield a duration of \SI[separate-uncertainty=true]{570(172)}{\atto\second} from Eq.~\eqref{BranchingRatio}. This lies quantitatively in the same range of the estimated duration. Hence, conversely, using measured field enhancement factors for different wavelengths, we should be able to estimate the interaction duration for each wavelength from Eq.~\eqref{BranchingRatio} in future work. 

In conclusion, we have demonstrated visibility spectroscopy of quantum pathway interference in multi-photon photoemission at a tungsten needle tip, driven by two-color laser fields. A roughly constant visibility of 90\% $\pm$ 5\% over an almost octave-spanning fundamental laser wavelength range from \SI{1180}{\nano\meter} to \SI{2000}{\nano\meter} was observed, suggesting a universal quantum pathway interference. This observation is well explained by the proposed theoretical model, which strongly supports the coherent emission physics to be strictly limited to the replacement of two fundamental photons with one second harmonic photon --- despite various other possible processes. Furthermore, the reported visibility scaling with respect to the laser intensity ratio is perfectly matched by our model. We expect this fundamental quantum path interference model to hold as long as the fundamental frequency is lower than the plasma frequency of the material and at least two fundamental photons are required to drive photoemission. Even more importantly, we expect this model to be applicable to virtually any driving frequency and any material. In particular, we envision that two-color visibility spectroscopy will offer a powerful tool for gaining deep insights in multi-photon processes and photoemission dynamics in materials as different as individual atoms, molecules, clusters and nanomaterials as well as extended surfaces. Examples include access to the quantum phases of the electronic states involved, the identification of resonant intermediate states, and measurement of decoherence effects due to environmental couplings (see supplement for more details). 

\begin{acknowledgments}
The authors thank Michael Kr\"{u}ger for insightful discussion. This work was funded by the Gordon and Betty Moore Foundation (GBMF) through Grant No. GBMF4744  ''Accelerator on a Chip International Program-ACHIP'', BMBF via 05K16WEC and 05K16RDB, ERC Consolidator Grant ''Near Field Atto'', DFG SPP 1840 and DFG SFB 953.
\end{acknowledgments}

\end{document}


\title{Supplemental Material: Quantum interference visibility spectroscopy in two-color photoemission from tungsten needle tips}

\author{Ang Li}
\email[]{ang.li@fau.de}
\affiliation{Department of Physics, Friedrich-Alexander Universit\" at Erlangen-N\" urnberg (FAU), \\ Staudtstra\ss e 1, 91058 Erlangen, Germany}

\author{Yiming Pan}
\affiliation{Physics Department and Solid-State Institute, Technion, \\ Haifa 32000, Israel}

\author{Philip Dienstbier}
\affiliation{Department of Physics, Friedrich-Alexander Universit\" at Erlangen-N\" urnberg (FAU), \\ Staudtstra\ss e 1, 91058 Erlangen, Germany}

\author{Peter Hommelhoff}
\email[]{peter.hommelhoff@fau.de}
\affiliation{Department of Physics, Friedrich-Alexander Universit\" at Erlangen-N\" urnberg (FAU), \\ Staudtstra\ss e 1, 91058 Erlangen, Germany}

\date{\today}


\maketitle
\section{Derivation of the proportionality factors $\alpha$, $\beta$ and branching ratio $\Gamma$}
The final state population via pathway I and II is described in Eq. (3), from which we can further derive the visibility from its definition Eq. (1) in this paper as, 
\begin{align}
\label{Visibility}
    V & = \frac{2\frac{\left| J_{\mathrm{-1}}\mathrm{(\eta_{\mathrm{2\omega}})} J_{\mathrm{-2}}\mathrm{(\eta_{\mathrm{\omega}})} \right|}{\left| J_{\mathrm{-4}}\mathrm{(\eta_{\mathrm{\omega}})} \right|}}{1+ \left( \frac{\left| J_{\mathrm{-1}}\mathrm{(\eta_{\mathrm{2\omega}})} J_{\mathrm{-2}}\mathrm{(\eta_{\mathrm{\omega}})} \right|}{\left| J_{\mathrm{-4}}\mathrm{(\eta_{\mathrm{\omega}})} \right|} \right)^2} ,  \\
    \label{visibility2}
      & = \frac{2 \sqrt{\Gamma \Tilde{I}}}{1+\Gamma\Tilde{I}}. 
\end{align}
where $J_{\mathrm{n}}$ are the $n$th-order Bessel functions of the first kind, $n$ corresponds to the order of the multi-photon process. It is obvious, from Eq.~\eqref{Visibility} and Eq.~\eqref{visibility2} we have the following relation, 
\begin{align}
    \sqrt{\Gamma \Tilde{I}} & = \frac{\left| J_{\mathrm{-1}}\mathrm{(\eta_{\mathrm{2\omega}})} J_{\mathrm{-2}}\mathrm{(\eta_{\mathrm{\omega}})} \right|}{\left| J_{\mathrm{-4}}\mathrm{(\eta_{\mathrm{\omega}})} \right|} , \\
    \label{JJ/J}
    & = \frac{\eta_{\mathrm{2\omega}}}{\frac{1}{24} \eta_{\mathrm{2\omega}}^2}. 
\end{align}
Here, we applied the asymptotic form of the Bessel function of the $n$-th order $J_{\mathrm{n}}(\eta_{\mathrm{\omega}})=(-2)^n/(-n!\cdot\eta_{\mathrm{\omega}}^n)$ with the effective exchanged photon number $\eta_{\mathrm{\omega}}= \frac{e k \Delta t g_{\mathrm{\omega}}}{m} F$. For simplicity reasons, we substitute the parameters $\frac{e k \Delta t g_{\mathrm{\omega}}}{m} $ with $\chi_{\omega}$. Therefore, we obtain $\eta_{\mathrm{\omega}}=\chi_{\mathrm{\omega}}F_{\mathrm{\omega}}=\sqrt{\chi_{\mathrm{\omega}}^2 \frac{2 I_{\mathrm{\omega}}}{\epsilon_0 c}}$, where $\epsilon_0$ is the vacuum permittivity, $c$ is the speed of light in vacuum and $I_\omega$ is the laser intensity. Now, Eq.~\eqref{JJ/J} can be further written as, 
\begin{align}
    \label{JJ/J2}
    \Gamma \Tilde{I} & =  \frac{\frac{2\chi_{\mathrm{2\omega}}}{\epsilon_0 c} I_{\mathrm{2 \omega}}}{\left( \frac{\chi_{\mathrm{\omega}}}{12\epsilon_0 c} I_{\mathrm{\omega}} \right)^2} ,  \\
    \label{JJ/J3}
    & = \frac{\beta I_{\mathrm{2 \omega}}}{\alpha^2 I_{\mathrm{\omega}}^2}, 
\end{align}
where $\alpha$ and $\beta$ are the proportionality factors for the absorption of $\omega$ and $2\omega$ photons, respectively. By comparing Eq.~\eqref{JJ/J2} with Eq.~\eqref{JJ/J3} we get the values of these factors as, 
\begin{align}
    \label{alpha}
   \alpha & = \frac{\chi_{\mathrm{2 \omega}}}{12\epsilon_0 c} = \frac{ e^2 k_{\mathrm{F}}^2 \Delta t^2}{12 m^2 \epsilon_0 c } \cdot \frac{ \xi_{\mathrm{\omega}}^2}{\omega^2},  \\
    \label{beta}
    \beta & = \frac{2\chi_{\mathrm{\omega}}}{\epsilon_0 c} = \frac{2 e^2 k_{\mathrm{F}}^2 \Delta t^2}{ m^2 \epsilon_0 c } \cdot \frac{ \xi_{\mathrm{2\omega}}^2}{\omega^2}.
\end{align}
Hence, the material branching ratio $\Gamma$ can be easily obtained as, 
\begin{equation}
\label{BranchingRatio}
    \Gamma = \frac{\beta}{\alpha^2}=\frac{72 m^2 \epsilon_0 c}{ e^2 k_{\mathrm{F}}^2 \Delta t^2} \cdot \frac{\omega^2 \xi_{\mathrm{2\omega}}^2}{\xi_{\mathrm{\omega}}^4} . 
\end{equation}

\section{Field enhancement factor simulation and material branching ratio calculation}

\begin{figure}[H]
\centering
\includegraphics[width=0.6\textwidth]{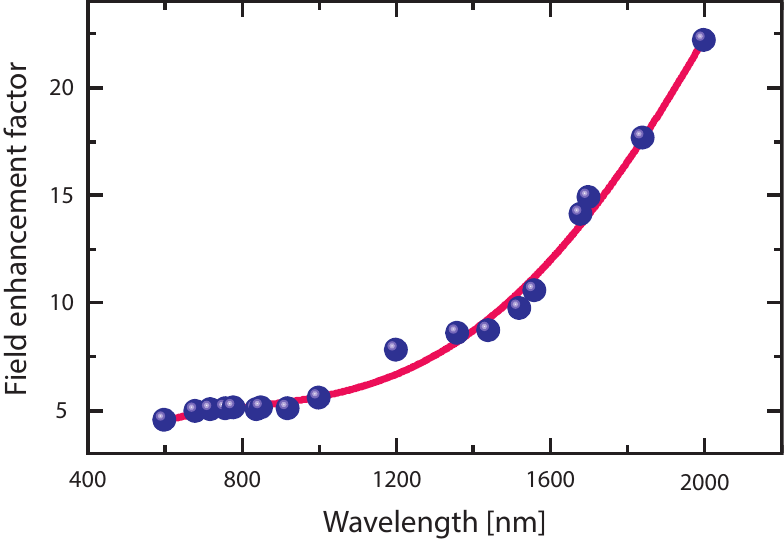}
\caption{\label{fig:FEF} Field enhancement factor calculated as a function of wavelength. The red solid line shows a polynomial fit of the fourth order to the FDTD simulation data (blue spheres). }
\end{figure}

The field enhancement factors $\xi_{\omega}$ and $\xi_{2\omega}$ in Eq. (6) are obtained from a finite-difference time-domain (FDTD) simulation \cite{Thomas2015}. In the simulation a tungsten needle tip with tip radius of \SI{14}{\nano\meter} is illuminated by a laser beam perpendicular to the tip axis. The laser beam is linearly polarized parallel to the tip axis and has a focal spot size obtained from experiment. Here, the wavelength of the laser beam is varied and the peak field at the surface of the tip apex is compared with the incident field to obtain the field enhancement factor, shown in Fig.~\ref{fig:FEF}. The simulation results (blue spheres) are fitted with a polynomial function of the fourth order (red solid line) to interpolate the data. 

\begin{figure}[!ht]
\centering
\includegraphics[width=0.6\textwidth]{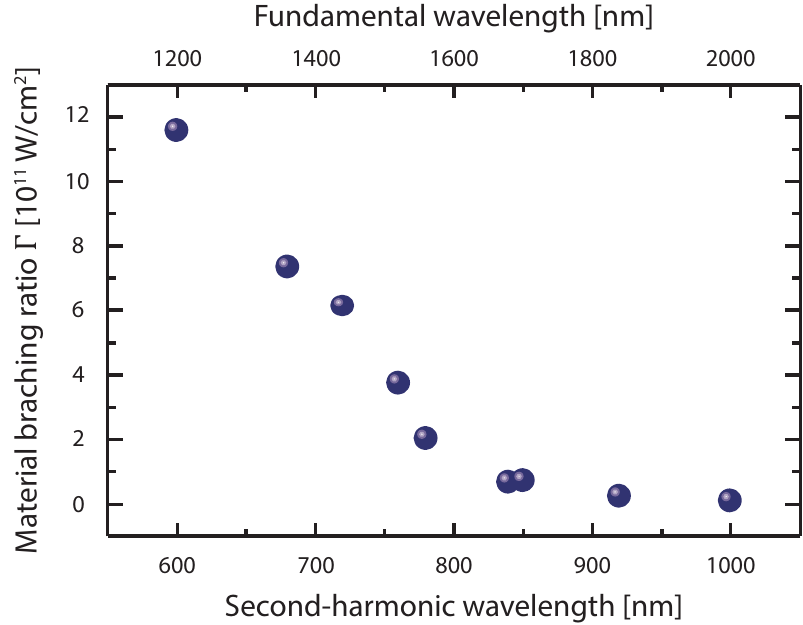}
\caption{\label{fig:coefficient} Material branching ratio $\Gamma$ calculated with various wavelengths. }
\end{figure}

The material branching ratio $\Gamma$ calculated from Eq.~\eqref{BranchingRatio} with the simulated field enhancement factors are shown in Fig.~\ref{fig:coefficient}. Thus, we can calculate the values of $\Gamma$ used in the visibility scaling (blue curves in Fig 3 (a) - (c) in this paper) as $\Gamma_{\mathrm{720nm}}=$ \SI{6.1e11}{\watt\per\centi\meter\squared}, $\Gamma_{\mathrm{850nm}}=$ \SI{7.2e10}{\watt\per\centi\meter\squared} and $\Gamma_{\mathrm{780nm}}=$ \SI{2.0e11}{\watt\per\centi\meter\squared}. Moreover, from the fitted curves in Fig. 3 (a) - (c) in this paper, we obtain the material branching ratios $\Gamma$ for the measured wavelength pairs, that yield $\Gamma_{\mathrm{720nm}}=$ \SI[separate-uncertainty = true]{2.2(3)e11}{\watt\per\centi\meter\squared}, $\Gamma_{\mathrm{850nm}}=$ \SI[separate-uncertainty = true]{1.3(1)e11}{\watt\per\centi\meter\squared} and $\Gamma_{\mathrm{780nm}}=$ \SI[separate-uncertainty = true]{1.3(1)e11}{\watt\per\centi\meter\squared}. These values of $\Gamma$ show an excellent agreement for different wavelength and intensities as they are shown in Fig. 3 (a) - (c) in the this paper. 

\newpage
\section{Potential applications of two-color quantum interference visibility spectroscopy} 

The advantage of the two-color ($\omega$-2$\omega$) quantum visibility spectroscopy with precisely controlled time delay, intensity ratio, and wavelengths is two-fold. First, it allows us to reveal the quantum phase information of the atomic or molecular states involved in the quantum interference associated with the two pathways of the multiphoton absorption processes. Once the pathway interference is observed via the visibility, we can extract the quantum phase via scanning the time-delay and intensity admixture ratio, see Eq. (3) in this paper. Hence, this method allows direct access to the quantum phase of an important state, which might be otherwise hard to attain. Second, the two-color inference relates in general to the linear response (single photon process -O(2$\omega$)) and the second order response (two-photon process -O$(\omega)^2$). It can thus help to target and identify a or several resonant intermediate states in all kind of systems by scanning the employed wavelengths. Thus, we expect that this method will facilitate studying multiphoton responses, nonlinear transitions and relaxations (also spin-forbidden, for example), ionization, and dissociation for diverse non-equilibrium quantum systems and platforms, such as Rydberg atoms, quantum gases, vacancy centers in solids and other artificial atoms, trapped ions, and in the realm of nuclear magnetic resonance. The photocurrent quantum interference visibility spectroscopy can be easily combined with widely-used two-color pump-probe experiments. Still, we should note that the quantum interference visibility spectroscopy we studied here is limited by the non-negligible two-photon nonlinear processes (otherwise, no interference can show up) and environment-induced decoherence (the interference information is suppressed) – but the latter point might even allow deep insights to non-equilibrium processes if probed with variable time delays.